\documentclass[english,aps,prstper,reprint,showpacs,titlepage,longbibliography]{revtex4-2}   

\usepackage[T1]{fontenc}	
\usepackage[latin9]{inputenc}	
\usepackage{geometry}		
\usepackage{graphicx}
\usepackage[above,below]{placeins}	
\usepackage{times}

\usepackage{hyperref}  
\hypersetup{colorlinks=true,urlcolor=blue,citecolor=blue,linkcolor=blue}   
\usepackage{enumitem}          
\setlist{nosep}                 

\renewenvironment{quote}{
   \par\addvspace{0mm}
   \list{}{
     \leftmargin 5mm
     \rightmargin 5mm
     \itshape
   }
   \item\relax
}{\par\addvspace{0mm}\endlist}

\begin{document}

\begin{titlepage}

\title{All aboard! Challenges and successes in professional development for physics lab TAs}
\author{Danny Doucette}
\author{Russell Clark}
\author{Chandralekha Singh}
\affiliation{Department of Physics and Astronomy, University of Pittsburgh, Pittsburgh, PA, USA 15260}

\begin{abstract}
At large research universities in the USA, introductory physics labs are often run by graduate student teaching assistants (TAs). Thus, efforts to reform introductory labs should address the need for effective and relevant TA professional development. We developed and implemented a research-based professional development program that focuses on preparing TAs to effectively support inquiry-based learning in the lab. We identify positive effects by examining three possible ways in which the professional development might have impacted TAs and their work. First, we examine lab TAs' written reflections to understand the effect of the program on TAs' ways of thinking about student learning. Second, we observe and categorize TA-student interactions in the lab in order to investigate whether TA behaviors are changing after the professional development. Third, we examine students' attitudes toward experimental science and present one example case in which students' attitudes improve for those TAs who `buy in' to the professional development. Our results suggest lab TA professional development may have a tangible positive impact on TA performance and student learning.\clearpage
\end{abstract}

\maketitle
\end{titlepage}

\section{Introduction and Framework}

At large research universities where introductory physics labs are often taught by graduate student teaching assistants (TAs), these TAs may not receive adequate professional development for their work~\cite{GilreathTATraining, SinghCategorizeProblems, LinTAProblems, GoodTAViews, MarshmanContrastingGrading, MariesTUGK, MariesFCI, KarimCSEM, WilcoxRIOT, DeBeckTABeliefs}. We are seeking to understand and develop a model for how TAs can most effectively help support student learning in physics labs, and what type of professional development can most effectively help TAs facilitate lab activities so that all students can learn. Our goal is to establish a lab TA professional development program that teaches TAs how students learn in an experimental physics setting, empowers lab TAs to improve their instructional practice, and ultimately produces long-overdue learning and attitudinal outcomes~\cite{HolmesWiemanLabsNoEffect,ECLASSsummary,ZwicklLabEpistemology,Hu2018ECLASS,QuinnGenderLabs,KoenigLabs} for all students in an inclusive and equitable learning environment.

Past work has identified several important elements of effective TA professional development~\cite{RecruitingRetainingFutureEducators}, including adapting good ideas to one's local context~\cite{JossemTAResourceLetter}, establishing a purposeful community of practice~\cite{HolmesTATraining}, and focusing on the development of the TA's beliefs and identity as an educator~\cite{, GoertzenTABeliefs, GrettonTATraining}. Another key issue is `buy-in': TAs who do not believe in the value of the learning activity will tend to implement it with low fidelity~\cite{WilcoxTABuyin}, which generally negatively impacts student learning~\cite{KoenigTutorialEffectiveness}. Achieving buy-in is a complex effort that depends on the context of the professional development and a variety of social cues that, when effective, work together to help TAs come to value the planned learning activities~\cite{GoertzenTABuyin}.

We adopt the cognitive apprenticeship model~\cite{CognitiveApprenticeship} as a theoretical framework for understanding both the evolution of TA learning in our professional development program and the nature of student learning in the labs. In this view, we recognize that learning requires three stages: modeling, scaffolding and coaching, and weaning. First, TAs in our training program need to witness explicit modeling of the desired outcomes. Second, learning requires careful scaffolding that supports evidence-based active engagement, and so TAs should be coached and provided guidance and support in learning how to provide this type of assistance. And third, this scaffolding and support should be gradually removed, giving TAs opportunities to practice independently. Thus, we understand TA professional development to be effective if our TAs learn about and employ effective strategies for supporting student learning, and if our students demonstrate elevated educational outcomes as a result of their TA's support.

\section{TA Professional Development}

At our large research university, approximately 33 introductory physics lab sessions are run every year. In each `cookbook'-style lab, up to 24 students work at lab benches, typically in pairs. Labs meet for 3 hours, once per week, for one semester. Although assignments vary, typically 5 to 12 TAs are assigned to teach the introductory labs in any given semester. For many, it is their first time teaching a lab course.

Some professional development is already provided for graduate student TAs in the physics department at our university. However, all of this professional development assumes that the TAs will be small-class lecturers or recitation leaders, rather than lab instructors. This professional development includes a day-long workshop that focuses on TAs' formal responsibilities as employees, a 3-hour workshop designed to give them strategies to effectively lead recitations, and a one-credit course that teaches about effective pedagogy and affords practice with recitation-style work.

Thus, while graduate student TAs receive a variety of instructional professional development at our university, there was nothing designed specifically for lab TAs. This was a concern because some of the skills, attitudes, and approaches that are required for TAs to be effective in lab settings are not the same as those needed in recitations. Additionally, after their first semester in graduate school, most TAs are unlikely to receive any formal professional development. To rectify these deficiencies, we designed a new professional development program for lab TAs that started in the fall 2018 semester and was replicated in the spring 2019 and summer 2019 semesters. These sessions involved students using `cookbook'-style labs that are, in general, not very effective in promoting student learning~\cite{RealTimePhysics}. Nonetheless, while we transition to an inquiry-based curriculum, we wanted to investigate the impact of professional development on lab effectiveness given this constrained setting. 

In our program~\cite{LabTATraining}, lab TAs meet weekly on Friday afternoons, including the Friday before the first week of classes, to prepare for the forthcoming week's lab, and to learn and practice lab-relevant pedagogy. The meetings wrap up mid-semester in order to reinforce the idea that the professional development program is a scaffold from which the TAs can be weaned, in line with our cognitive apprenticeship model, and that we expect TAs to continue developing as educators beyond our professional development program. The program was developed via extensive discussions and iterations between the three authors, all experienced physics educators.

We identified early on that motivation would be key to making this professional development program effective. Following the interest framework of Hidi and Renninger~\cite{HidiRenningerMotivation}, we developed learning activities for TAs that would trigger situational interest in their work as TAs, reinforce that interest through meaningful social reflections, and consequently establish sustained individual interest in becoming effective instructors in their labs and beyond.

Situational interest reinforcement happens at the start of our weekly meetings, when all the TAs share a student interaction from the past week they found surprising, concerning, or encouraging. These reflections provoke cross-discussion in which TAs celebrate their progress as educators, reaffirm shared commitments to helping students learn meaningfully, or brainstorm approaches to uncommon problems. These discussions are moderated by the training leader, using standard methods for establishing and maintaining positive interactivity~\cite{CohenDesigningGroupwork}. At all times, we focus on giving TAs opportunities to speak out in order to develop their confidence as educators.

Most of the lab TA meeting is dedicated to one or two relevant learning activities, which are designed to promote TAs' individual interest as educators. The learning activities are intended to help TAs directly improve their skills at working with students, better understand the nature of student learning~\cite{PiagetOriginsIntelligence}, and develop both proficiency with the apparatus and increased levels of motivation to support students.

One learning activity we employed is role-playing student-TA interactions around key points in the lab. For this, the trainer sets up `sabotaged' experiments in which the apparatus is miscalibrated, the analysis is incorrectly done, or a similar common issue. The scaffolding in this experiment allows TAs to practice interactions that support evidence-based active learning. After role-playing through an interaction, the TAs and trainer debrief and move to another experiment. Two other approaches we used occasionally are demonstrations, which serve as models for TAs to replicate, and conducting carefully-moderated whole-group discussions. Some sample learning activities are listed in Table \ref{Activities}.

\begin{table}[htbp]
\caption{Sample of learning activities in the lab TA professional development program.\label{Activities}}
\begin{ruledtabular}
\begin{tabular}{llr}
Activity & Type & Weeks \\
\hline
Icebreaker & demonstration & first \\
Reflections & group discussion & all \\
'Sabotaged' experiments & role-playing & most \\
Equitable learning environments & group discussion & third\\
Teaching about the nature of science & group discussion & fifth\\
\end{tabular}
\end{ruledtabular}
\end{table}

\vspace{-0.2cm}

\section{Methodology and Results}

We assess the impact of our lab TA professional development program by analyzing three different examples of ways in which the training program might have impacted TAs or students. For clarity, the methods and results of each of these three examples are presented together. We begin by asking whether the program changes how TAs think about their work in supporting student learning. Next, we investigate the behavior of TAs as they work in their lab sections to understand if the professional development has changed the nature of their interactions with students. Last, we ask whether the program has a `second-order' effect by improving students' learning outcomes.


\textbf{A. TAs' attitudes toward student learning.} To assess the direct effect of the professional development program, lab TAs were asked to write short responses to variations on the question, ``How will you help students have effective learning experiences?'' at the start and end of the program. In total, 13 responses were collected at the start, and 11 at the end, of the fall and spring programs.

The responses show a marked transformation in how the TAs viewed their role in helping students to learn. At the beginning of the professional development program, most of the TAs described their role in terms of a traditional transmission model of education~\cite{DeweyExperienceAndEducation}, as these representative excerpts illustrate:

\begin{quote}
    I will explain to them which equipment corresponds to which concept. Then they can build connections of the physics concepts to the lab.
\end{quote}

\begin{quote}
    [I will] add in some physical explanation into the demo at the start of lab... a feeling on physics will build up subconsciously after they leave the lab.
\end{quote}

The use of the verb `explain' is abundant in these early responses, as the TAs view themselves as either telling students about physics concepts or clarifying the procedure for the lab-work. Other responses emphasize the importance of good lecture structure, clear expectations for lab report formatting, and creating a ``relaxing'' environment for the students. Given that most of the TAs have experienced a traditional, transmission-based style of education -- and are continuing to experience that model as graduate students -- it is not surprising that they rely on the transmission model of learning to frame their work. Likewise, it makes sense that the TAs view their role as helping to make the lab easier for students, as that is likely how they viewed their own TAs during their recent undergraduate experiences. 

By the end of the the professional development program, however, the emphasis shifts and most of the TAs' responses celebrate students figuring things out on their own and with their lab partners, as seen in these typical quotations:

\begin{quote}
    I would encourage a student... to collaborate with peers, ask themselves more rigorous questions, etc.
\end{quote}

\begin{quote}
    I asked her to think of the problem in a physical sense, instead of plugging in given equations. She actually came up with the right answer... by herself.
\end{quote}

These final reflections indicate that after the professional development the TAs have generally discarded the transmission model in favor of a student-centered view of learning, which was one of our goals. `Explain' is no longer used, and the responses tend to center the student's experience instead of the TA's work. Other responses emphasize the importance of encouraging positive collaboration between students and explain techniques the TAs use for supporting student meaning-making without directly supplying information.

It is not possible to conclusively determine what caused the shift in TAs' views about learning. Was it the professional development program? Their experiences as a TA in the lab? Something else? However, by comparing experienced TAs who have taught the lab before (in semesters before the professional development program) with newer TAs who have never taught labs before, it seems likely that the common factor -- the lab TA professional development -- played at least some role in their movement from a transmission to a constructivist view of learning.

\begin{figure}[b]
\includegraphics[width=\columnwidth]{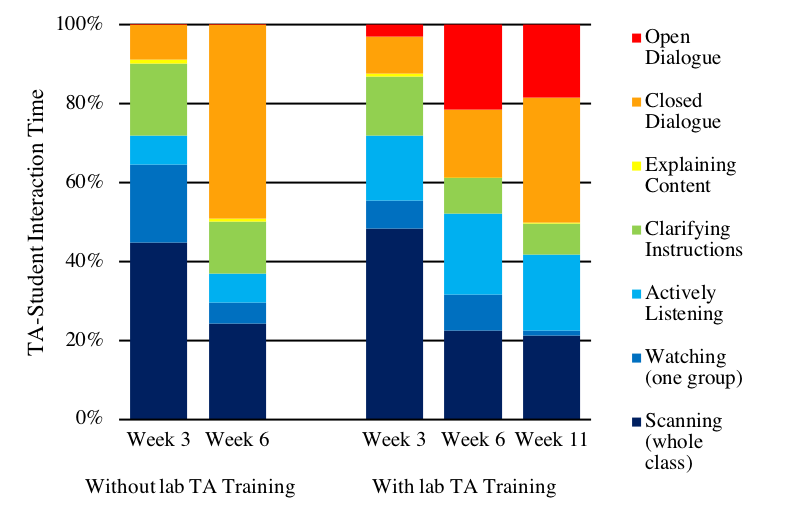}
\caption{RIOT data from a semester before the introduction of a lab TA professional development program compared with data from the fall 2018 semester, when the program was implemented. The vertical axis shows the fraction of time spent in each of 7 different types of interaction, averaged over the 6 and 7 TAs observed during that semester.\label{RIOTplot}}
\end{figure}


\textbf{B. TA-Student Interactions.} Since one goal of the lab TA professional development is to get TAs to help students frame and answer their own questions, rather than just offering advice and explanations, we hypothesized that there would be a change in the nature of student-TA interactions after introducing the program. To measure the extent of these changes, we used the Real-time Instructor Observing Tool (RIOT) \cite{RIOT} to categorize these interactions for the same group of TAs.

The RIOT allows an observer to continuously categorize the types of activities undertaken by an instructor, in this case the lab TA. When the instructor switches from one type of activity to another, the observer records the nature of the new activity along with a timestamp. Other than infrequent occasions when the TA would be briefly checking personal notes or be out of the room, the seven categories shown in Fig. \ref{RIOTplot} capture the full breadth of TA activities during the lab sessions that we observed. The data presented in Fig. \ref{RIOTplot} depicts interactions for 13 TAs over 99 hours of instruction. We chose to observe during weeks 3 and 6 of our 13-week semester because the lab interactions should have reached a `steady state' and because the lab-work for those weeks was typical, not excessively intricate, and didn't require that the lights be turned off. We also observed during week 11 for the TAs that received the professional development to explore if the effect of the training diminished after the meetings wrapped up in week 6. Complete definitions and examples of the seven types of interactions we observed are available in Ref. \cite{RIOT}.

We adopted slightly different labels for some categories for clarity, but use the same definitions and meanings for these interaction types. For example, we use `Actively Listening' (to question) to emphasize that the TA is engaging with non-verbal listening cues, unlike in the case of `Watching'. Likewise, `Explaining Content' and `Clarifying Instructions' are forms of "talking at students"~\cite{RIOT}. We exclude several interaction types that are not relevant to our labs and the relatively rare cases when the TA is not interacting with students.

Two specific examples of these categories will be relevant to our analysis. `Open Dialogue' is an interaction in which a student is contributing more than half the words (or ideas) to a conversation and is actively developing an understanding of physics or lab ideas, while the TA plays a supporting role by asking prompting questions or helping students to frame their ideas, as opposed to `Closed Dialogue' which is TA-dominated. `Actively Listening' means that the TA is near a group of students and showing an active interest in their discussion through non-verbal cues such as establishing eye contact, body positioning, or gestures, but isn't participating in the conversation as a contributor. 

Two differences stand out in Fig. \ref{RIOTplot}. First, the amount of time that TAs devote to `Open Dialogue' is larger for the TAs who have completed the professional development program, and it seems to increase while the TAs are engaged in the program. This behavior aligns with our goals and with the responses we discussed in section IIIA: the TAs are more likely to let students lead their conversations, prompting rather than telling, and helping the students to formulate and answer their own questions.

Second, the TAs who have taken the professional development program devote approximately twice as much time to `Active Listening' as those who haven't taken the program. In practice, this manifests as the TAs being more willing to engage with students and offer non-verbal support and attention as the students complete their lab-work. In our observations, this increased level of `Active Listening' appears to be aligned with TAs being increasingly comfortable interacting with students when they can position themselves as guides rather than feeling compelled to adopt the role of an all-knowing expert. In other words, the TAs were comfortable simply being there and talking with the students, and didn't feel the need to adopt a `hide or provide' behavior, in which they only approached student groups if they felt they had some information to share.

Teaching is certainly a complex endeavor, and no one interaction type is necessarily better than any other in all circumstances. Good educators typically use a combination of many types of interaction \cite{UsingRIOT}. Overall, though, we find some evidence that the balance of interaction types after the professional development program is better-aligned to the goal of supporting student-led active learning for TAs who have taken the program.


\textbf{C. Student Outcomes.} While both TA attitudes and interactions show improvements, the fundamental goal in our new TA professional development program is to improve student learning in the labs. Thus, in order to ascertain the impact of our lab TA program on student learning, we need to compare students whose TAs did deliver the learning support we designed in our professional develop program with those students whose TAs did not deliver this support. Here we discuss it in relation to the lab TA professional development module related to the nature of science (NoS).

The nature of science is a set of beliefs about the epistemology of science, i.e.: principles that we might identify as fundamental characteristics of Western scientific work. These include such ideas as `theories require evidence', `science makes predictions', and `scientists seek to avoid bias'~\cite{NatureOfScience}.

We observed that for the fall 2018 implementation, our professional development module for the nature of science had an abnormally low level of engagement from our TAs. During the program, most of the TAs merely went through the motions, and during lab observations later in the semester we saw no evidence that they used the proposed strategies in their labs. However, two TAs clearly and unambiguously bucked the trend: they engaged in a lengthy discussion about the value of explicit instruction on the nature of science that went beyond the training session, and we observed them using our strategies in their lab sessions on multiple occasions. Thus, by comparing the 74 students who received this NoS treatment from their TA to the 207 who did not, we estimate the impact of having a TA engaged with NoS instruction.

\begin{figure}[b]
\includegraphics[width=\columnwidth]{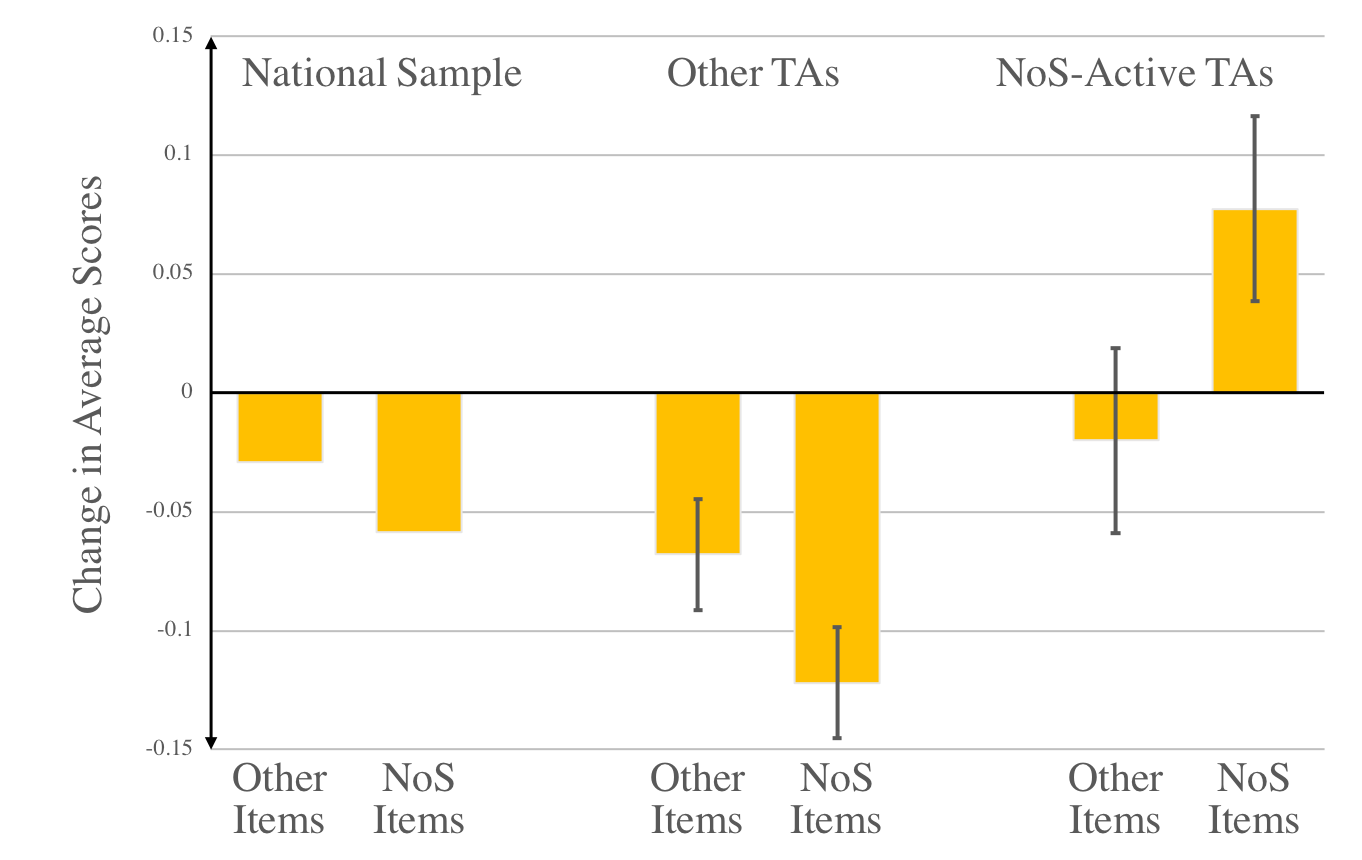}
\caption{Change in E-CLASS item scores between pre- and post-instruction assessments, averaged over the 24 and 6 E-CLASS items identified as either not, or belonging to, a cluster related to the nature of science. TAs who were active in discussing NoS during the program and who explicitly talked about it in their labs are compared to those who did not. National comparison sample from \cite{ECLASSValidation}.\label{ECLASSitems}}
\end{figure}

As a dependent variable, we rely on a categorization of E-CLASS items~\cite{ECLASSValidation} that identified a cluster of 6 items as relevant to NoS in our context~\cite{DoucetteLabOne}. E-CLASS scores indicate the degree to which a student agrees with expert-like attitudes toward 30 items related to experimental science. As shown in Fig.~\ref{ECLASSitems}, when we consider the `Other TAs' that did not engage with the value of teaching NoS, the average change in score for the 6 NoS-related items is more negative than for each of the 24 other E-CLASS items, indicating that our students typically regress more on the nature of science items than on the other items in E-CLASS. And while the decrease in our setting is somewhat more negative than in the national sample from \cite{ECLASSValidation}, a similar pattern appears, suggesting that these 6 NoS-related items are particularly vulnerable.

For the TAs who have adopted NoS instruction, we observe a decrease on the 24 non-NoS items that is comparable to the decrease seen by the other TAs, but a sizable increase on the 6 NoS-related items. Using a mixed-effect ANCOVA model controlling for overall pre-test E-CLASS score, and with satisfactory normality and homogeneity of variance, we find a statistically significant difference between our two TA groups and the two item categories ($F(1,279)=11.5, p<0.001$).

We note that the lab-work in this study employed a `cookbook'-style approach that will be replaced with an evidence-based active learning approach \cite{RealTimePhysics} starting in fall 2019. We expect that overall E-CLASS score improvements will require both that TAs are effectively trained (and adopt the strategies learned in this professional development) and that inquiry-based learning activities are in place.

We can draw two conclusions from this result. First, students attitudes about experimental physics, as measured by at least some of the E-CLASS items, can be influenced by the interactions and learning that are offered by graduate student TAs in the introductory lab. Second, TA buy-in for particular instructional strategies is essential in order for this to happen.

\section{Discussion and Implications}

\vspace{-0.1in}

At three levels of analysis, we find evidence that an effective lab TA professional development program has the potential to positively impact the work undertaken by TAs and the learning of their students. After the professional development program, lab TAs demonstrated a shift in how they viewed their role as instructors and how they thought about the nature of student learning. It appears that the lab TAs who completed the program were also more likely to `walk the walk,' interacting with students in ways that are commensurate with what we sought to teach them about supporting active engagement learning. And when lab TAs took up the new approach to teaching a topic, such as explicitly engaging in discussions about the nature of science, student attitudes corresponding to that topic became more expert-like.

The nature of the work done by lab TAs, the activities involved in our program, and the improvements noted above act together to shine light on the importance of specialized professional development for TAs who are working in lab settings. While there is overlap with how we might train TAs to support evidence-based active engagement learning, the need to develop mastery at troubleshooting apparatus and to address issues such as the nature of science means that lab TAs should be receiving dedicated professional development in order to be effective in their work.

As seen in the case of our nature of science module, some improvement is still needed for our professional development program. TA `buy-in' remains a vital issue, especially for the NoS module, and as we go forward we will continue working to improve the social and structural factors that can drive TA buy-in~\cite{GoertzenTABuyin}. However, even with an imperfect professional development program and `cookbook'-style labs, and through the inevitable noise of implementation, these results suggests that lab TA professional development has the potential to have a positive impact on TA performance and effectiveness.


\bibliography{the}

\end{document}